\documentclass{aa}
\usepackage{graphicx}
\usepackage{natbib}
\begin{document}

   \title{X-ray Detections of Two Young Bona-Fide Brown Dwarfs \thanks{Based on observations obtained with the ESA \emph{XMM-Newton} Satellite, reported in the The First XMM-Newton Serendipitous Source Catalogue,  XMM-Newton Survey Science Centre (SSC), 2003}}

   \author{H. Bouy \inst{1,2,3} }

   \offprints{H. Bouy}

   \institute{Max Planck f\"ur Extraterrestrische Phyisk, Giessenbachstra\ss e, D-85748 Garching bei M\"unchen, Germany\\
             \email{hbouy@mpe.mpg.de}
	     \and Laboratoire d'Astrophysique de l'Observatoire de Grenoble, 414 rue de la piscine, F-38400 Saint Martin d'H\`ere, France\\
	     \and European Southern Observatory, Karl Schwarzschildtstra\ss e 2, D-85748 Garching bei M\"unchen, Germany\\
	                 }

   \date{Received 06/01/2004 ; accepted 04/05/2004}

   \abstract{I report here the detection of two bona-fide brown dwarfs by \emph{XMM-Newton}: \object{[GY92]~141} in the $\rho$-Ophiuchus star forming region and \object{DENIS~J155601-233809} in the Upper Scorpius OB association. The two objects have been detected with luminosities of $L_{X}$=8.35$\pm$2.86$\times$10$^{28}$~erg s$^{-1}$ and $L_{X}$=6.54$\pm$1.35$\times$10$^{28}$~erg s$^{-1}$ respectively, corresponding to luminosity ratios of log($L_{X}/L_{bol}$)= -2.07 and log($L_{X}/L_{bol}$)= -2.69. The two sources were close to the limit of detection of the instruments at only 2$\sim$3-$\sigma$ above the background level, and no significant flare or variation could be detected during the 48.3~ks and 33.9~ks observations. [GY92]~141 had already been observed 10 months earlier with \emph{Chandra} by \citet{2001ApJ...563..361I} with a luminosity $\sim$14 times fainter than the one I report here, meaning that the X-ray emission of this object is strongly variable.
}

\authorrunning{Bouy}   

\maketitle

\markboth{X-Ray Detections of Two Young Bona-Fide Brown Dwarfs}{Bouy H.}

%
%________________________________________________________________

\section{Introduction}
Brown dwarfs are very low mass objects unable to sustain stable nuclear reactions. Although these objects are extremely cool and their atmospheres fully convective, they appear to be able to display some X-ray coronal activity. Several surveys in the field or in star forming regions using \emph{ROSAT} \citep{2002A&A...391.1025M, 2000A&AS..146..323N, 1999A&A...343..883N, 1998ApJ...504..461F}, \emph{XMM-Newton} \citep{2003PASJ...55..653, 2003IAUS..211..443S, 2002NewA....7..595M} or \emph{Chandra} \citep{2003IAUS..211..447W, 2003ApJ...587L..51T, 2002AJ....123.1613P, 2002AAS...201.4608A, 2001ApJ...563..361I, 2000ApJ...538L.141R, 2000ApJ...533..372F} lead to the identification of several ultracool and brown dwarfs as faint X-ray emitters, with typical luminosities of $L_{X}\sim$10$^{27\sim28}$erg s$^{-1}$, and some of them displaying strong variability. These observations suggest that there is apparently no drop in the X-ray luminosity at the substellar boundary. Although these X-ray features are likely to have a magnetic origin, the standard ($\alpha$--$\omega$) dynamo cannot be at work because of  the absence of a radiative core and the fully convective nature of these objects. \citet{1999A&A...346..922K} suggested that the magnetic field could be generated by a so-called $\alpha^{2}$--dynamo effect, where the magnetic field is produced by the action of the Coriolis force on the convective zones of the atmosphere. The X-ray emission might then be correlated to the rotation, and therefore the age, of the object, which is still not clear as shown by the X-ray detection of old late-M dwarfs in the field \citep{2002NewA....7..595M, 2000ApJ...538L.141R, 2000ApJ...533..372F, 1998ApJ...504..461F}. Finally, since no abrupt changes are observed at the convective transition, the $\alpha^{2}$--dynamo would probably be present in the convective zones of more massive stars. The study of the X-ray activity of ultracool and brown dwarfs will therefore bring extremely important results not only on their own characteristics but also on the properties of more massive objects. 

In this paper, I will present the detection of two young brown dwarfs with the ESA \emph{XMM Newton} satellite. In the first section, I will present these two objects and their optical and/or infrared properties; in section \ref{obs} I will describe the observations, then in section \ref{data} I will explain how I processed and analyzed the data and finally in section \ref{analysis} I will analyze the results.

\section{Two Young Brown Dwarfs \label{two_bd}}
The two objects have been confirmed as brown dwarfs and as members of the Upper Scorpius and $\rho$-Oph star forming regions by spectroscopic measurements. In this section, I will summarize the properties of these objects as reported in the literature.

\subsection{DENIS~J155601-233809}
DENIS~J155601-233809 has been reported as a M6.5 dwarf and as a member of the Upper Scorpius OB association by \citet{martin_usco} on the basis of spectroscopic measurements. Proper motion measurements confirm that the object belongs to the association (Bouy et al., in prep.). \citet{martin_usco} report a clear H$\alpha$ emission with an equivalent width of EW(H$\alpha$)=-20$\pm$3~\AA, not strong enough to indicate significant accretion, but indicating that DENIS-P~J155601-233809 is chromosphericaly active (see e.g \citet{Barrado...Martin} or \citet{2003ApJ...592..282J} for a discussion on the H$\alpha$ emission as an indicator of accretion and activity).

DENIS~J155601-233809 was also reported in the 2MASS catalogue as \object{2MASS~J155601-233808}. Table \ref{targets} gives an overview of the photometric and astrometric properties of DENIS~J155601-233809. DENIS and 2MASS photometry (see Table \ref{targets}) indicate a $I-J$ colour of 2.46$\pm$0.10~mag. If we assume an intrinsic colour for M6.5 dwarfs of $I-J$=2.5--2.7~mag as reported by \citet{2002AJ....124.1170D}, the measured $I-J$ colour corresponds to an almost null extinction A$_{V}$$\sim$0~mag. Using the bolometric correction $B_{K}$=2.95$\pm$0.15~mag evaluated for M6.5 dwarfs by \citet{2002ApJ...564..452L}, the 2MASS apparent magnitude $m_{K}$=12.81$\pm$0.03~mag, and assuming a distance of 145$\pm$2~pc as measured with \emph{Hipparcos} by \citet{1999AJ....117..354D}, I arrive at log $L/L_{\sun}$=-2.08$\pm$0.05.

\subsection{[GY92]~141}
[GY92]~141 (hereafter GY~141) was reported by \citet{1997ApJ...489L.165L} as a M8.0 dwarf and as member of the $\rho$-Oph star forming region (\object{$\rho$--Oph~162349.8-242601}) using spectroscopic measurements. On 1997 April 14-15th they report a relatively strong H$\alpha$ emission with an equivalent width of EW(H$\alpha$)$\sim$-60~\AA. On 2002 October 31 \citet{2002ApJ...578L.141J} report an H$\alpha$ emission with an equivalent width of EW(H$\alpha$)=-13.4$\pm$0.2~\AA. The variability and the strength of this emission suggest chromospheric activity, but part of it could be due to accretion. The detection of mid-infrared excess emission by \citet{1998A&A...335..522C} with a magnitude of 12.25$\pm$0.3~mag at 4.5~$\mu$m in the LW1 filter of the ISOCAM indeed indicates the presence of circumstellar material. \citet{1997ApJ...489L.165L} estimated a mass of 0.01--0.06~M$_{\sun}$ using the evolutionary tracks of \citet{1997ApJ...491..856B} and \citet{1997A&A...327.1054B}, and \citet{2003ApJ...593.1074G} estimated a mass of 0.024$\pm$0.015~M$_{\sun}$. GY~141 is therefore clearly substellar. 

GY~141 has already been observed twice in the X-rays, first in September 1994 with \emph{ROSAT} by \citet{1999A&A...343..883N} who did not detect it with an upper limit on the luminosity at log $L_{X}\le$27.86, and second by \citet{2001ApJ...563..361I} with \emph{Chandra} in May 2000, who detected it with a luminosity $L_{X}=$0.25$\times$10$^{28}$~erg/s and log ($L_{X}/L_{bol}$)=-3.6. The source was too faint to allow them to make spectroscopy. They also estimated the chance of coincidence of background source to fall in their 9\arcsec\, beam to be 0.6\%--1\%.

GY~141 was also reported in the DENIS catalogue as \object{DENIS~J162651.3-243242.9} and in the 2MASS catalogue as \object{2MASS~J16265113-243242}. Table \ref{targets} gives an overview of the photometric and astrometric properties of this object as reported in these two catalogues.

\section{Observations \label{obs}}

The two sources have been reported in the XMM-Newton Serendipitous Source Catalogue \citep{XMM_Cat}. By doing a cross-matching between a list of currently know brown dwarfs and brown dwarf candidates with the catalogue, I found that two X-ray sources had been reported at the position of these two objects. I retrieved the corresponding data from the XMM-Newton Public Archive (program 0112380101, P.I. Turner, and program 0111120201, P.I. Watson) for further analysis. 

\subsection{DENIS~J155601-233809}
\object{DENIS~J155601-233809} has been observed on 2000 August 26th for 48309~s. The pointing was made at R.A= 15\degr56\arcmin25.00\arcsec \, and Dec=-23h37\arcmin47.02\arcsec, therefore 331\arcsec\, away from DENIS~J155601-233809.  The observation has been made in prime full window mode with the medium filter, which prevents optical contamination from point sources as bright as $m_{v}$=8$\sim$10~mag. Both MOS and PN detectors have a 27\arcmin\, diameter field of view. DENIS~J155601-233809 appears only on the EPIC PN and MOS2 images. It is reported in the XMM-Newton Serendipitous Source Catalogue as  \object{1XMM~J155601.1-233809}. Figure \ref{position} shows the combination of the EPIC PN, MOS1 and MOS2 images and the position of DENIS~J155601-233809. For more details on \emph{XMM-Newton} and its instruments, please refer to the XMM-Newton Users' Handbook \citep{XMM_UH}.

The X-Ray source lies 1\farcs0 away from the DENIS and the 2MASS coordinates, therefore well within the absolute astrometric uncertainties of the pointing (4\arcsec \, in average, according to the XMM-Newton Users' Handbook). No other optical/near infrared counterpart can be found in the DENIS or 2MASS catalogues within an area of $\sim$20\arcsec\, around this position, which makes it very likely that the faint X-ray source is associated to the DENIS object. Moreover and as will be discussed in Section \ref{data}, the associated flux is consistent with that of a brown dwarf. The corrected detection likelihood value reported in the catalogue is 44.8, with a quality flag of 4, ensuring that the probability that the detection is real is very high. \footnote{According to the User Guide to the Catalogue, \emph{``the likelihood value stands for the detection likelihood of the source, $L=-ln(p)$, where $p$ is the probability the detection occurring by chance. A value of $\sim$7 corresponds roughly to one false detection per exposure. A summary quality flag of 4 indicate that the detection is good (0=bad; 1-3=suspect, 4=good)''. For more details please refer to the User Guide to the Catalogue.}}

\subsection{GY~141 \label{gy141}}
\begin{figure*}
   \centering
   \includegraphics[width=\textwidth]{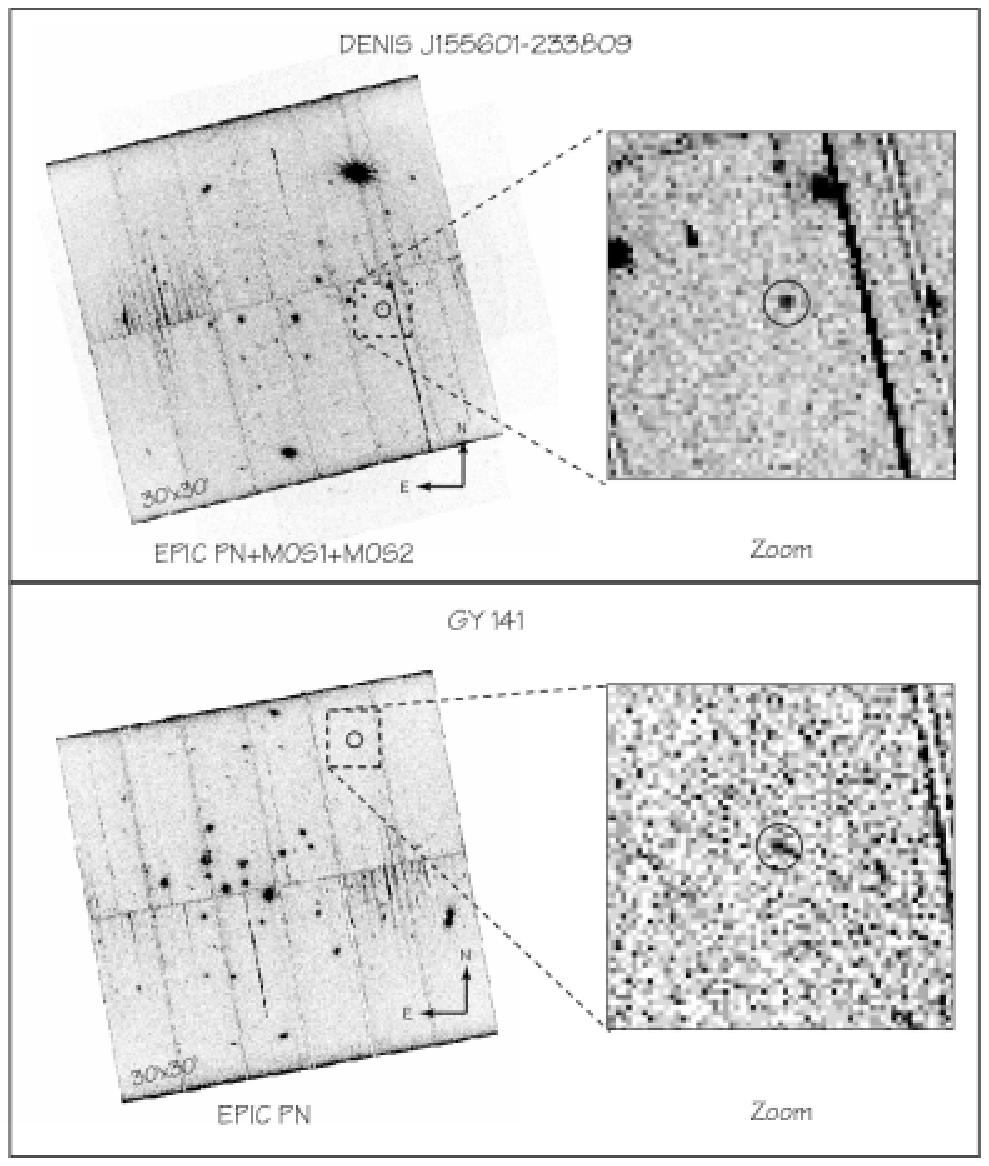}
   \caption{This figure shows the \emph{XMM-Newton} images where the two objects have been detected. In the case of DENIS~J155601-233809 it shows the combination of the images obtained with the 3 instruments (EPIC MOS1, MOS2 and PN cameras) and in the case of GY~141 only the EPIC PN image.}
   \label{position}
   \end{figure*}

\object{GY~141} has been observed on 2001 February 19th for 33908~s. The pointing was made on the core of the $\rho$--Oph star forming region, 722\arcsec \, away from GY~141. The observation has also been made in prime full window mode with the medium filter. Figure \ref{position} shows the EPIC PN image and the position of GY~141 in the field. GY~141 was not detected in the EPIC MOS2 camera. This source is reported in the XMM-Newton Serendipitous Source Catalogue as  \object{1XMM~J162651.3-243242}. 
The catalogue does not report any detection in the EPIC PN camera, although the source appears clearly in it (see Figure \ref{position}). In both EPIC PN and MOS1, the source is very faint, which might explain why the pipeline used to build the catalogue missed it in the EPIC PN image. The corrected detection likelihood value reported in the catalogue for the MOS1 detection is 9.0, with a quality flag of 4. According to the User Guide to the Catalogue, sources with likelihood greater than $\sim$8 and quality flag of 4 are reliable. GY~141 is just above this limit, which probably explains why it was not detected by the pipeline in the other EPIC detectors. Using an aperture of 320 pixels, I measure 92.6 counts in the EPIC PN image (corrected for background using a standard \emph{phot} sky algorithm). The corresponding rate is 0.00273$\pm$00092~count/s, which is consistent with the value reported in the catalogue for the EPIC MOS1 camera (see Table \ref{x_ray_counts}). The detection in two of the three detectors increases the likelihood that the detection is real, and the chances of coincidence are very low. On the basis of all these considerations, I will consider for the rest of this paper that the detection is real, although the possibility that it is spurious cannot be ruled out.

The X-ray source lies less than 0\farcs5 away from the DENIS and 2MASS coordinates, and once again no optical/near infrared counterpart can be found in the DENIS and 2MASS catalogues in an area of $\sim$25\arcsec\, around this position, which makes it very likely that the faint X-ray source is associated with the brown dwarf. As will be discussed in Section \ref{data}, the associated flux is also consistent with that of a brown dwarf, but higher than the one reported by \citet{2001ApJ...563..361I} with \emph{Chandra}.

   \begin{table}
         \caption[]{Astrometric and photometric measurements available on the two objects}
         \label{targets}
     \begin{tabular}{@{} lllll @{}}
           \hline
 R.A.\footnotemark[1] &      Dec.\footnotemark[1]   &  Filter   & Mag.   &  Source \\
            \hline
\multicolumn{5}{c}{DENIS~J155601-233809} \\
\hline
15 56 01.04     &  -23 38 08.1  &   J     &  13.86$\pm$0.03  &  2MASS \\
                &               &   H     &  13.24$\pm$0.02  &  2MASS \\
                &               &   K     &  12.81$\pm$0.03  &  2MASS \\
15 56 01.04     &  -23 38 08.1  &   I     &  16.32$\pm$0.07  &  DENIS \\
                &               &   J     &  13.96$\pm$0.11  &  DENIS \\
                &               &   K     &  12.85$\pm$0.14  &  DENIS \\
\hline
\multicolumn{5}{c}{GY~141} \\
\hline
16 26 51.28     &  -24 32 42    &   J     &  15.30$\pm$0.04  &  2MASS \\
                &               &   H     &  14.47$\pm$0.05  &  2MASS \\
                &               &   K     &  13.89$\pm$0.06  &  2MASS \\
16 26 51.31     &  -24 32 42.9  &   I     &  18.42$\pm$0.18  &  DENIS \\
                &               &   J     &  15.33$\pm$0.15  &  DENIS \\
                &               &   K     &  13.47$\pm$0.18  &  DENIS \\
            \hline
\thanks{\footnotemark[1] J2000}
     \end{tabular}

   \end{table}

\section{Data Processing \label{data}}

\begin{figure*}
   \centering
   \includegraphics[width=\textwidth]{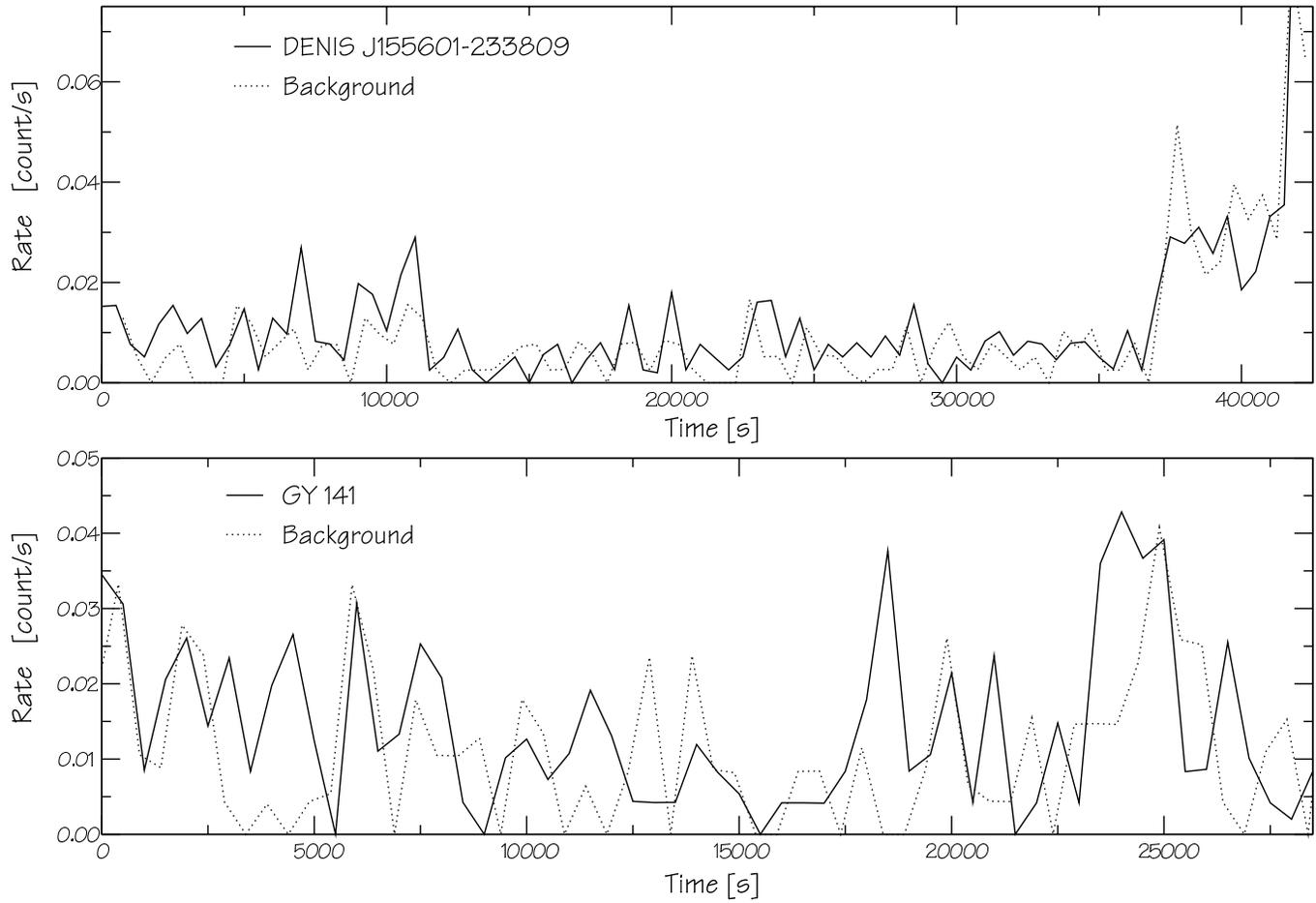}
   \caption{Light-curves of DENIS~J155601-233809, GY~141, and the background around their positions. The size of the time bins is 500~s. In both case the source and the background have very similar amplitudes and variations, indicating that the sources are at the limit of detection. Figure \ref{position} also shows that the two sources are very faint.}
   \label{lc}
   \end{figure*}

Since the source were not centered in the field of view, I had to apply a correction for vignetting using the recommended \emph{evigweight} task of the XMM-Newton Science Analysis System software \citep{XMM_SAS}. Once this correction was applied, I extracted the light curve of each object in a circular area of 400 pixels around the source, and the light-curve of their respective background in an empty annulus region between 450 and 602 pixels around the source, therefore with the same surface in order to make direct comparison. Figure \ref{lc} shows the light-curves obtained. It appears clearly that the sources are very faint and not much brighter than the background. Neither timing nor spectral analysis can therefore reasonably be performed on these data, and I was only able to get flux measurements. 

The fluxes reported in the XMM-Newton Serendipitous Source Catalogue were computed for $N_{H}$=3.0$\times$10$^{20}$~cm$^{2}$, and assuming a spectral model of a power-law with a slope of 1.7. For more details about the construction of the catalogue, please refer to the corresponding User Guide \footnote{XMM-Newton Survey Science Centre Consortium (http://xmmssc-www.star.le.ac.uk/)}. This spectral model being not the most adapted to this kind of sources, I decided to recompute the fluxes using the count rates reported in the catalogue (see Table \ref{x_ray_counts}, except for the PN values of GY~141, measured as explained in section \ref{gy141}) and the recommended \emph{PIMMS} and \emph{nH} softwares\footnote{\emph{PIMMS} and \emph{nH} are distributed by the NASA's HEARSAC (http://heasarc.gsfc.nasa.gov/)}.

Table \ref{x-ray_lum} gives the results obtained assuming a thin thermal plasma with a temperature $kT$ varying from 0.5 to 2.5, a distance of 145~pc for both objects as measured by \citet{1999AJ....117..354D}, and bolometric luminosities estimated as explained in Section \ref{two_bd}.

\begin{table}
              \caption[]{X-ray Count Rates\footnotemark[1]  }
         \label{x_ray_counts}
\begin{center}
     \begin{tabular}{lc}
           \hline
Detector & Count Rates     \\
         &  [10$^{-3}\cdot$counts$\cdot$s$^{-1}$]  \\
\hline
\multicolumn{2}{c}{DENIS~J155601-233809} \\
\hline
MOS1     &  0.59$\pm$0.20  \\
MOS2     &  not detected   \\
PN       &  3.25$\pm$0.49  \\
MOS1+PN\footnotemark[2] &  2.48$\pm$0.40 \\
\hline
\multicolumn{2}{c}{GY~141} \\
\hline
MOS1     &  1.66$\pm$0.46  \\
MOS2     &  not detected   \\
PN       &  2.73$\pm$0.92   \\
MOS1+PN\footnotemark[2] &  2.37$\pm$0.77 \\
\hline
     \end{tabular}
\end{center}
\thanks{\footnotemark[1] Source: XMM-Newton Serendipitous Source Catalogue, except EPIC PN value for GY~141}\\
\thanks{\footnotemark[2] mean total count rate of the detections weighted by the errors}\\
\end{table}

\begin{table*}
\begin{center}
     \caption[]{X-ray flux and luminosities (in the energy range 0.2--12~keV)}
     \label{x-ray_lum}
     \begin{tabular}{lcccc}
\hline
Detector &  $kT$\footnotemark[1]  &    Flux           &    L$_{X}$      &  log $L_{X}/L_{bol}$ \\
         &       [keV]   & [10$^{-15}$ergs$\cdot$cm$^{-2}\cdot$s$^{-1}$]   & [10$^{28}$ergs$\cdot$s$^{-1}$] & \\
\hline
\multicolumn{5}{c}{DENIS~J155601-233809} \\
\hline
MOS1     & 1.0 &  7.87$\pm$2.67    &  1.98$\pm$0.67  &  -3.21   \\
         & 1.5 &  12.48$\pm$4.24   &  3.14$\pm$1.06  &  -3.00   \\
         & 2.0 &  17.57$\pm$5.96   &  4.42$\pm$1.50  &  -2.86   \\
         & 2.5 &  22.32$\pm$7.58   &  5.61$\pm$1.91  &  -2.75   \\
PN       & 1.0 &  15.26$\pm$2.32   &  3.84$\pm$0.58  &  -2.92   \\
         & 1.5 &  23.37$\pm$3.55   &  5.88$\pm$0.89  &  -2.73   \\
         & 2.0 &  31.28$\pm$4.75   &  7.87$\pm$1.19  &  -2.61   \\
         & 2.5 &  38.06$\pm$5.77   &  9.57$\pm$1.45  &  -2.52   \\

MOS1+PN\footnotemark[2] &  1.0  &  12.08$\pm$2.49    & 3.04$\pm$0.62  &  -3.02  \\
         & 1.5 &  18.88$\pm$3.89   &  4.75$\pm$0.97  &  -2.83   \\
         & 2.0 &  25.95$\pm$5.36   &  6.54$\pm$1.35  &  -2.69   \\
         & 2.5 &  32.28$\pm$6.67   &  8.12$\pm$1.68  &  -2.59   \\
\hline
\multicolumn{5}{c}{GY~141} \\
\hline
MOS1     & 1.0 &  22.30$\pm$6.22   &  5.61$\pm$1.56  &  -2.25   \\
         & 1.5 &  35.23$\pm$9.82   &  8.86$\pm$2.47  &  -2.05   \\
         & 2.0 &  49.49$\pm$13.80  &  12.44$\pm$3.47 &  -1.90   \\
         & 2.5 &  62.82$\pm$17.52  &  15.80$\pm$4.41 &  -1.80   \\

PN       & 1.0 &  13.01$\pm$4.38   &  3.27$\pm$1.10  &  -2.49   \\
         & 1.5 &  19.81$\pm$6.67   &  4.98$\pm$1.68  &  -2.30   \\
         & 2.0 &  26.44$\pm$8.91   &  6.65$\pm$2.24  &  -2.18   \\
         & 2.5 &  32.12$\pm$10.83  &  8.08$\pm$2.72  &  -2.09   \\

MOS1+PN\footnotemark[2] &  1.0  &  16.09$\pm$5.30    & 4.05$\pm$1.33  &  -2.39  \\
         & 1.5 &  24.67$\pm$8.24   &  6.21$\pm$2.08   &  -2.20   \\
         & 2.0 &  33.22$\pm$11.35  &  8.35$\pm$2.86  &  -2.07   \\
         & 2.5 &  40.61$\pm$14.17  &  10.21$\pm$3.57  &  -1.99   \\
\hline

     \end{tabular}
\end{center}
\thanks{\footnotemark[1] temperature of the thermal plasma spectral model used to convert from count rate to flux} \\
\thanks{\footnotemark[2] mean total flux/luminosity of the detections weighted by the errors}

   \end{table*}

\section{Data analysis \label{analysis}}

Since no spectral analysis could be performed to determine which spectral model best describes the sources, I will make the analysis using the results obtained for a thermal plasma with $kT$=2.0~keV as shown in Table \ref{x-ray_lum}.

\subsection{DENIS~J155601-233809}
As shown in Figure \ref{logL_spt}, DENIS~J155601-233809 has an X-ray luminosity very similar to that of other late type M-dwarfs in star forming regions and the field. It belongs to the strongest X-ray emitters among that class of objects. It is important to note that no detailed timing analysis can be done because of the very low flux (see Figure \ref{lc}), and that more sensitive observations are required in order to know if the observed emission is quiescent or related to some flare events.

\subsection{GY~141}
The X-ray activity of GY~141 shows that this object has a strong coronal activity. In order to compare the different measurements available, I used the \emph{PIMMS} software to convert the \emph{Chandra}  luminosity (measured in the energy range 0.5--9.0~keV) and the \emph{ROSAT} upper limit (measured in the energy range 0.1--2.4~keV) to the energy range of the \emph{XMM-Newton} observations (0.2--12.0~keV). The results are summarized in Table \ref{comparison}, and show that the flux of GY~141 increased by a factor of $\sim$14 between the \emph{Chandra} and the \emph{XMM-Newton} observations. The X-ray emission is therefore highly variable, and GY~141 would not have been detected in the 33.9~ks \emph{XMM-Newton} frame if its emission would have been at the same level than in the 100.6~ks \emph{Chandra} observation. The present \emph{XMM-Newton} measurement is close to the limit of detection of the \emph{ROSAT} observation.

\begin{table}
              \caption[]{X-ray Luminosities (0.2--12~keV) of GY~141 at different epochs}
         \label{comparison}
     \begin{tabular}{lccc}
           \hline
Instrument  &  Date Obs.   &  $L_{X}$   &  log $L_{X}/L_{bol}$ \\
            &              &  [10$^{28}$ergs$\cdot$s$^{-1}$] & \\
\hline
ROSAT\footnotemark[1]       & 09/1994      &  $\le$13.00      &  $\le$-1.9  \\
Chandra     & 05/2000      &  0.60      &  -3.22      \\
XMM-Newton  & 02/2001      &  8.35      &  -2.07      \\
\hline
     \end{tabular}
\thanks{\footnotemark[1] GY~141 was not detected. The corresponding value is an upper limit.}\\
\thanks{Remarks: Values computed assuming a thermal plasma spectral model with $kT$=2~keV.}
\end{table}

\begin{figure*}
   \centering
   \includegraphics[width=\textwidth]{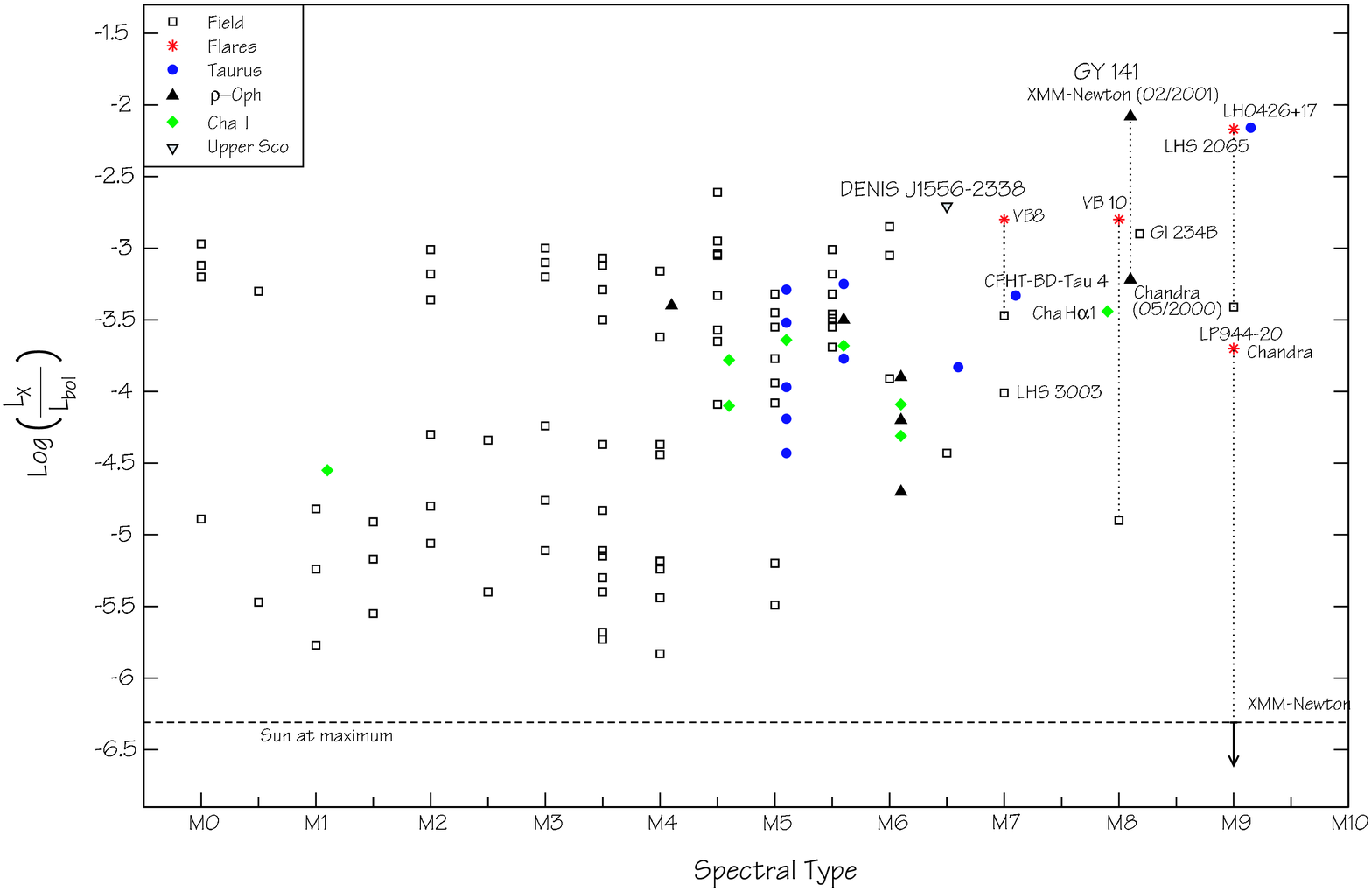}
   \caption{Distribution of X-ray luminosity of M dwarfs as a function of spectral type. The open squares represent the values obtained for field M-dwarfs. The circles, diamonds, triangles up and down denote young objects in star-forming regions. The stars represent the flaring objects at their maximum with the corresponding value or upper limit on the quiescent emission. While DENIS~J155601-233809 seems to have a normal X-ray luminosity in comparison with other M6$\sim$M7 dwarfs, GY~141 shows a strong variability between the \emph{Chandra} and \emph{XMM-Newton} observations.  In order to avoid confusion, some object's spectral types have been shifted by 0.1 subclass. Values from \citet{2002NewA....7..595M} and references therein, except the quiescent value for \object{VB~10}, from \citet{2003ApJ...594..982F}.}
   \label{logL_spt}
   \end{figure*}

As shown in Figure \ref{logL_spt}, GY~141 has an X-ray luminosity among the strongest for late type objects, and displays a strong variability between two observations separated by a period of $\sim$10 months. \citet{1999A&A...343..883N} did not detect it with \emph{ROSAT} in 1994, but their limit of detection was higher than the two values obtained with \emph{XMM-Newton} and \emph{Chandra}. Its X-ray luminosity increased by a factor of $\sim$14 between the \emph{Chandra} and the \emph{XMM-Newton} observations. Although the light curve of the emission detected  with \emph{XMM-Newton} suggests that there was no significant variation during the 34~ks \emph{XMM-Newton} observation (see Figure \ref{lc}), the shortness of the observation (34~ks) and the faintness of the detection does not allow to conclude that it was not a flare-like event. Further observations should be made in order to confirm that the observed variability occurs on long ($\sim$months) or short ($\sim$10~ks) time-scales. This issue is very important since the only few sources known to show variability are flaring objects with periods of 10$\sim$50~ks and, except for the old field brown dwarfs \object{LP944-20} and \object{VB~10}, with amplitudes less than a factor of 10 \citep{2003PASJ...55..653}. If confirmed by further observations, a strong variability over a long time-scale would probably mean that two types of activity, governed by different mechanisms, can occur in the corona of late type objects. Recently, \citet{2002A&A...382L...9S} reported a significant variation of the quiescent X-ray emission of the M9 dwarf \object{LHS~2065} after a flare. Such a phenomenon might also be at work and explain the variation observed for GY~141 between the \emph{Chandra} and \emph{XMM-Newton} observations. Finally, since it has been shown that the object is surrounded by a disk, part of its emission could also come from the circumstellar material interacting with the magnetic field, following a scenario of emission as suggested by \citet{2000ApJ...532.1097M}.

\section{Conclusions}
I report here the detection with \emph{XMM-Newton} of two young bona-fide brown dwarfs. DENIS~J155601-233809 displays an X-ray emission similar to other young objects of the same age and spectral class, with log($L_{X}/L_{bol}$)= -2.69. I did not observe any significant variability during the 48.3~ks observation, but the faintness of the source did not allow me to perform precise timing analysis. More sensitive observations are required to know if this emission is quiescent or associated to a flare-like event. GY~141 has the highest X-ray luminosity reported up to date for that class of object, with  log($L_{X}/L_{bol}$)= -2.07. Its luminosity has increased by a factor of 14 since the previous \emph{Chandra} observation. No significant flare-like event could be detected in the 33.9~ks \emph{XMM-Newton} observation. In that case again the faintness of the source did not allow me to perform precise timing analysis, and more sensitive observations are also required in order to know if this emission is due to a flare or quiescent. Follow-up observations of this object should allow us to know if this variability occurs on long ($\sim$months) or short ($\sim$10~ks) time-scales.

\begin{acknowledgements}
I would like to thank Elisa Constantini for her precious and generous help regarding the processing of XMM-Newton data. I would like to thank Eduardo Mart\'\i n and Xavier Delfosse for providing me the list of Upper Scorpius brown dwarfs prior to publication, and J\'er\^ome Bouvier and Wolfgang Brandner for their kind support. 

This research has made use of the XMM-Newton Public Archive and of the XMM-Newton Serendipitous Source Catalogue \citep{XMM_Cat}. This research has also made use of the  VizieR on-line Service at Centre de Donn\'es astronomiques de Strasbourg, France \citep{Vizier}. This publication makes use of data products from the Two Micron All Sky Survey (2MASS) and DEep Near Infrared Survey (DENIS). The Two Mass Survey is a joint project of the University of Massachusetts and the Infrared Processing and Analysis Center/California Institute of Technology, funded by the National Aeronautics and Space Administration and the National Science Foundation. The DENIS project has been partly funded by the SCIENCE and the HCM plans of the European Commission under grants CT920791 and CT940627. It is supported by INSU, MEN and CNRS in France, by the State of Baden-Württemberg in Germany, by DGICYT in Spain, by CNR in Italy, by FFwFBWF in Austria, by FAPESP in Brazil, by OTKA grants F-4239 and F-013990 in Hungary, and by the ESO C\&EE grant A-04-046. 

\end{acknowledgements}

\bibliographystyle{aa}
%\bibliography{mybiblio}

\begin{thebibliography}{34}
\expandafter\ifx\csname natexlab\endcsname\relax\def\natexlab#1{#1}\fi

\bibitem[{{Adams} {et~al.}(2002){Adams}, {Wolk}, {Walter}, {Jeffries}, \&
  {Naylor}}]{2002AAS...201.4608A}
{Adams}, N., {Wolk}, S., {Walter}, F.~M., {Jeffries}, R., \& {Naylor}, T. 2002,
  Bulletin of the American Astronomical Society, 34, 1176

\bibitem[{{Baraffe} {et~al.}(1997){Baraffe}, {Chabrier}, {Allard}, \&
  {Hauschildt}}]{1997A&A...327.1054B}
{Baraffe}, I., {Chabrier}, G., {Allard}, F., \& {Hauschildt}, P.~H. 1997, \aap,
  327, 1054

\bibitem[{{Barrado y Navascues} \& {Mart\'\i n}(2003)}]{Barrado...Martin}
{Barrado y Navascues}, D. \& {Mart\'\i n}, E.~L. 2003, \aj, in press

\bibitem[{{Burrows} {et~al.}(1997){Burrows}, {Marley}, {Hubbard}, {Lunine},
  {Guillot}, {Saumon}, {Freedman}, {Sudarsky}, \&
  {Sharp}}]{1997ApJ...491..856B}
{Burrows}, A., {Marley}, M., {Hubbard}, W.~B., {et~al.} 1997, \apj, 491, 856

\bibitem[{{Comeron} {et~al.}(1998){Comeron}, {Rieke}, {Claes}, {Torra}, \&
  {Laureijs}}]{1998A&A...335..522C}
{Comeron}, F., {Rieke}, G.~H., {Claes}, P., {Torra}, J., \& {Laureijs}, R.~J.
  1998, \aap, 335, 522

\bibitem[{{Dahn} {et~al.}(2002){Dahn}, {Harris}, {Vrba}, {Guetter}, {Canzian},
  {Henden}, {Levine}, {Luginbuhl}, {Monet}, {Monet}, {Pier}, {Stone}, {Walker},
  {Burgasser}, {Gizis}, {Kirkpatrick}, {Liebert}, \&
  {Reid}}]{2002AJ....124.1170D}
{Dahn}, C.~C., {Harris}, H.~C., {Vrba}, F.~J., {et~al.} 2002, \aj, 124, 1170

\bibitem[{{de Zeeuw} {et~al.}(1999){de Zeeuw}, {Hoogerwerf}, {de Bruijne},
  {Brown}, \& {Blaauw}}]{1999AJ....117..354D}
{de Zeeuw}, P.~T., {Hoogerwerf}, R., {de Bruijne}, J.~H.~J., {Brown}, A.~G.~A.,
  \& {Blaauw}, A. 1999, \aj, 117, 354

\bibitem[{{Ehle} {et~al.}(2003){Ehle}, {Pollock}, {Talavera}, {Gabriel},
  {Chen}, {Verdugo}, {Ballet}, {Freyberg}, {Kirsch}, {Metcalfe}, {Osborne},
  {Pietsch}, {Saxton}, \& {Smith}}]{XMM_UH}
{Ehle}, M., {Pollock}, A.~M.~T., {Talavera}, A., {et~al.} 2003, XMM-Newton
  Users' Handbook, V2.1

\bibitem[{{Fleming}(1998)}]{1998ApJ...504..461F}
{Fleming}, T.~A. 1998, \apj, 504, 461

\bibitem[{{Fleming} {et~al.}(2003){Fleming}, {Giampapa}, \&
  {Garza}}]{2003ApJ...594..982F}
{Fleming}, T.~A., {Giampapa}, M.~S., \& {Garza}, D. 2003, \apj, 594, 982

\bibitem[{{Fleming} {et~al.}(2000){Fleming}, {Giampapa}, \&
  {Schmitt}}]{2000ApJ...533..372F}
{Fleming}, T.~A., {Giampapa}, M.~S., \& {Schmitt}, J.~H.~M.~M. 2000, \apj, 533,
  372

\bibitem[{{Gorlova} {et~al.}(2003){Gorlova}, {Meyer}, {Rieke}, \&
  {Liebert}}]{2003ApJ...593.1074G}
{Gorlova}, N.~I., {Meyer}, M.~R., {Rieke}, G.~H., \& {Liebert}, J. 2003, \apj,
  593, 1074

\bibitem[{{Imanishi} {et~al.}(2003){Imanishi}, {Nakajima}, {Tsujimoto},
  {Koyama}, \& {Tsuboi}}]{2003PASJ...55..653}
{Imanishi}, K., {Nakajima}, H., {Tsujimoto}, M., {Koyama}, K., \& {Tsuboi}, Y.
  2003, \pasj, 55, 653

\bibitem[{{Imanishi} {et~al.}(2001){Imanishi}, {Tsujimoto}, \&
  {Koyama}}]{2001ApJ...563..361I}
{Imanishi}, K., {Tsujimoto}, M., \& {Koyama}, K. 2001, \apj, 563, 361

\bibitem[{{Jayawardhana} {et~al.}(2002){Jayawardhana}, {Mohanty}, \&
  {Basri}}]{2002ApJ...578L.141J}
{Jayawardhana}, R., {Mohanty}, S., \& {Basri}, G. 2002, \apjl, 578, L141

\bibitem[{{Jayawardhana} {et~al.}(2003){Jayawardhana}, {Mohanty}, \&
  {Basri}}]{2003ApJ...592..282J}
{Jayawardhana}, R., {Mohanty}, S., \& {Basri}, G. 2003, \apj, 592, 282

\bibitem[{{K{\" u}ker} \& {R{\" u}diger}(1999)}]{1999A&A...346..922K}
{K{\" u}ker}, M. \& {R{\" u}diger}, G. 1999, \aap, 346, 922

\bibitem[{{Leggett} {et~al.}(2002){Leggett}, {Golimowski}, {Fan}, {Geballe},
  {Knapp}, {Brinkmann}, {Csabai}, {Gunn}, {Hawley}, {Henry}, {Hindsley},
  {Ivezi{\' c}}, {Lupton}, {Pier}, {Schneider}, {Smith}, {Strauss}, {Uomoto},
  \& {York}}]{2002ApJ...564..452L}
{Leggett}, S.~K., {Golimowski}, D.~A., {Fan}, X., {et~al.} 2002, \apj, 564, 452

\bibitem[{{Loiseau}(2003)}]{XMM_SAS}
{Loiseau}, N. 2003, XMM-Newton Science Analysis System Users' Guide, V2.1

\bibitem[{{Luhman} {et~al.}(1997){Luhman}, {Liebert}, \&
  {Rieke}}]{1997ApJ...489L.165L}
{Luhman}, K.~L., {Liebert}, J., \& {Rieke}, G.~H. 1997, \apjl, 489, L165+

\bibitem[{{Mart{\'{\i}}n} \& {Bouy}(2002)}]{2002NewA....7..595M}
{Mart{\'{\i}}n}, E.~L. \& {Bouy}, H. 2002, New Astronomy, 7, 595

\bibitem[{{Mart{\'{\i}}n} {et~al.}(2004){Mart{\'{\i}}n}, {Delfosse}, \&
  {Guieu}}]{martin_usco}
{Mart{\'{\i}}n}, E.~L., {Delfosse}, X., \& {Guieu}, S. 2004, Accepted for
  Publication in \aj

\bibitem[{{Mokler} \& {Stelzer}(2002)}]{2002A&A...391.1025M}
{Mokler}, F. \& {Stelzer}, B. 2002, \aap, 391, 1025

\bibitem[{{Montmerle} {et~al.}(2000){Montmerle}, {Grosso}, {Tsuboi}, \&
  {Koyama}}]{2000ApJ...532.1097M}
{Montmerle}, T., {Grosso}, N., {Tsuboi}, Y., \& {Koyama}, K. 2000, \apj, 532,
  1097

\bibitem[{{Neuh{\" a}user} {et~al.}(1999){Neuh{\" a}user}, {Brice{\~ n}o},
  {Comer{\' o}n}, {Hearty}, {Mart{\'{\i}}n}, {Schmitt}, {Stelzer}, {Supper},
  {Voges}, \& {Zinnecker}}]{1999A&A...343..883N}
{Neuh{\" a}user}, R., {Brice{\~ n}o}, C., {Comer{\' o}n}, F., {et~al.} 1999,
  \aap, 343, 883

\bibitem[{{Neuh{\" a}user} {et~al.}(2000){Neuh{\" a}user}, {Walter}, {Covino},
  {Alcal{\' a}}, {Wolk}, {Frink}, {Guillout}, {Sterzik}, \& {Comer{\'
  o}n}}]{2000A&AS..146..323N}
{Neuh{\" a}user}, R., {Walter}, F.~M., {Covino}, E., {et~al.} 2000, \aaps, 146,
  323

\bibitem[{{Ochsenbein} {et~al.}(2000){Ochsenbein}, {Bauer}, \&
  {Marcout}}]{Vizier}
{Ochsenbein}, F., {Bauer}, P., \& {Marcout}, J. 2000, \aaps, 143, 23

\bibitem[{{Preibisch} \& {Zinnecker}(2002)}]{2002AJ....123.1613P}
{Preibisch}, T. \& {Zinnecker}, H. 2002, \aj, 123, 1613

\bibitem[{{Rutledge} {et~al.}(2000){Rutledge}, {Basri}, {Mart{\'{\i}}n}, \&
  {Bildsten}}]{2000ApJ...538L.141R}
{Rutledge}, R.~E., {Basri}, G., {Mart{\'{\i}}n}, E.~L., \& {Bildsten}, L. 2000,
  \apjl, 538, L141

\bibitem[{{Schmitt} \& {Liefke}(2002)}]{2002A&A...382L...9S}
{Schmitt}, J.~H.~M.~M. \& {Liefke}, C. 2002, \aap, 382, L9

\bibitem[{{Stelzer} \& {Neuh{\" a}user}(2003)}]{2003IAUS..211..443S}
{Stelzer}, B. \& {Neuh{\" a}user}, R. 2003, in Brown Dwarfs, Proceedings of IAU
  Symposium 211, held 20-24 May 2002 at University of Hawai'i, Honolulu,
  Hawai'i. Edited by Eduardo Mart\'\i n., 443--+

\bibitem[{{Tsuboi} {et~al.}(2003){Tsuboi}, {Maeda}, {Feigelson}, {Garmire},
  {Chartas}, {Mori}, \& {Pravdo}}]{2003ApJ...587L..51T}
{Tsuboi}, Y., {Maeda}, Y., {Feigelson}, E.~D., {et~al.} 2003, \apjl, 587, L51

\bibitem[{{Wolk}(2003)}]{2003IAUS..211..447W}
{Wolk}, S.~J. 2003, in Brown Dwarfs, Proceedings of IAU Symposium 211, held
  20-24 May 2002 at University of Hawai'i, Honolulu, Hawai'i. Edited by Eduardo
  Mart\'\i n., 447--+

\bibitem[{{XMM-Newton Survey Science Centre}(2003)}]{XMM_Cat}
{XMM-Newton Survey Science Centre}. 2003, The First XMM-Newton Serendipitous
  Source Catalogue

\end{thebibliography}

\end{document}